\documentclass[conference]{IEEEtran}
\IEEEoverridecommandlockouts
\usepackage{cite}
\usepackage{amsmath,amssymb,amsfonts}
\usepackage{algorithmic}
\usepackage{graphicx}
\usepackage{textcomp}
\usepackage{xcolor}
\usepackage{hyperref}
\usepackage{tabularx}
\usepackage[utf8]{inputenc}
\usepackage{adjustbox}

\usepackage{float} % Add this line to use the [H] placement specifier
\def\BibTeX{{\rm B\kern-.05em{\sc i\kern-.025em b}\kern-.08em
    T\kern-.1667em\lower.7ex\hbox{E}\kern-.125emX}}
\begin{document}

\title{Mining Software Repositories for Expert Recommendation}
\author{\IEEEauthorblockN{Chad Marshall\IEEEauthorrefmark{1}, Andrew Barovic\IEEEauthorrefmark{2} and Armin Moin\IEEEauthorrefmark{3}}
\IEEEauthorblockA{\textit{Department of Computer Science, University of Colorado Colorado Springs}\\
United States \\
\IEEEauthorrefmark{1}cmarsha6@uccs.edu, \IEEEauthorrefmark{2}abarovi2@uccs.edu, \IEEEauthorrefmark{3}amoin@uccs.edu
}
}
\maketitle

\begin{abstract}
We propose an automated approach to bug assignment to developers in large open-source software projects. This way, we assist human bug triagers who are in charge of finding the best developer with the right level of expertise in a particular area to be assigned to a newly reported issue. Our approach is based on the history of software development as documented in the issue tracking systems. We deploy BERTopic and techniques from TopicMiner. Our approach works based on the bug reports' features, such as the corresponding products and components, as well as their priority and severity levels. We sort developers based on their experience with specific combinations of new reports. The evaluation is performed using Top-k accuracy, and the results are compared with the reported results in prior work, namely TopicMiner MTM, BUGZIE, Bug triaging via deep Reinforcement Learning BT-RL, and LDA-SVM. The evaluation data come from various Eclipse and Mozilla projects, such as JDT, Firefox, and Thunderbird. 
\end{abstract}


\section{Introduction}\label{introduction}
Large open-source projects offer an issue tracking system or open bug repository, where developers and users can report the software defects they find or any new feature requests they may have. These reports are called \textit{bug reports} or \textit{issues}. In some cases, developers can volunteer to work on the reported issues they find interesting or relevant to their field of expertise. Additionally, they sometimes report issues and assign them to themselves. However, in many cases, particularly in large open-source projects, a group of developers, called \textit{bug triagers}, decide who should process and fix a newly reported issue. In fact, \textit{bug triage} is the process of finding invalid and duplicate bug reports and assigning the valid and unique ones to the developers working on the project so they can fix and resolve them. 

Bug triagers typically take various factors, such as the developers' availability and areas of expertise, into account while assigning bug reports to developers. However, this is a tedious and costly task. Over the past decades, (semi-)automated approaches based on various Artificial Intelligence (AI) methods and techniques have been proposed to make the bug triage process, including bug assignment, more effective and efficient. This way, the total project costs will be reduced since the developers' (including bug triagers') time will be used more efficiently. Note that if a bug report is assigned to a developer with the wrong field of expertise or with a lack of proficiency in the matter, it has to be reassigned (i.e., \textit{tossed}) to another developer. This would be costly for the project since the developers' time is an invaluable resource for the project.

As explained in Section \ref{related-work}, prior works in the literature have studied the bug triage problem and proposed various AI-based approaches to address it, such as approaches based on Machine Learning (ML) \cite{Anvik+2006} (including graph learning \cite{Jeong+2009}) and Information Retrieval (IR) \cite{Sun+2011}. Most of them have used the natural language textual data in the bug reports as a crucial source of information. Hence, Natural Language Processing (NLP) plays a vital role. In general, automated bug triage can be considered as a sub-field of Mining Software Repositories (MSR).

In this paper, we propose a novel approach to automated bug assignment in open bug repositories. Our approach is based on recent advances in NLP. To this aim, we analyze the natural language textual information in the summary, description, and comments of existing bug reports for a specific project. First, we create a number of topic models and assign each report to one of them. This step is based on the work of Xia et al. \cite{Xia+2017}. Second, we train a text classifier for each topic that can predict the most suitable developer to take care of a specific bug report, given its summary and description (and, if available, comments).

The contribution of this paper is twofold: First, it proposes and implements a novel approach to automated bug assignment. This is validated by our experimental results using available data from open reference datasets deployed in the related work in the literature. Second, it provides an open-source prototype that enables other open-source projects using similar issue tracking systems to use the proposed approach, thus benefiting from effective and efficient automated bug assignment.

This paper is structured as follows: Section \ref{background} provides some background information about open bug repositories and NLP. Further, we review the literature in Section \ref{related-work}. In Section \ref{proposed-approach}, we propose our novel approach and report on our experimental results in Section \ref{experimental-results}. Moreover, Section \ref{discussion} discusses the results and points out potential threats to validity. Finally, we conclude and suggest future work in Section \ref{conclusion-future-work}.

\section{Background}\label{background}
This section discusses the necessary information for this paper and will help readers understand the approaches we will take.

\subsection{Open-bug repositories}
\textit{Open-bug repositories} allow the user and developer to interact with bug reports and other issues the user may experience when using the software. There is a diverse selection of \textit{bug reporting} software such as GitHub Issues\cite{fiechter2021visualizing} and Bugzilla \cite{1407819} to name a few. Bugzilla \cite{1407819} and GitHub Issues \cite{fiechter2021visualizing} are engineered to make it easy for the user to report bugs. Whenever a user reports a \textit{bug} on GitHub Issues \cite{fiechter2021visualizing}, the issue is posted online so anyone can see it; this allows other users to help resolve the issue if it is a common problem. Bugzilla \cite{1407819} only allows the user and developer to interact; this can help the developer get crucial information from the user to fix the issue. 

\subsubsection{Bug Reports}
\textit{Bug reports} are issues that users can report on software to a developer. The issues a user reports differ from systems and Operating systems (OS), so it is essential for \textit{bug reports} to track any valuable information to the developer. Bugzilla \cite{1407819} was one of the first commercially standard \textit{bug-tracking} software that added product and component sections to a \textit{bug report}. Adding product and component makes it easier for the developer to find the source of a bug; to track a bug, you must be able to locate it. Since the release of Bugzilla \cite{Anvik+2006}, the evolution of \textit{bug reports} from software like Jira and GitHub Issues \cite{fiechter2021visualizing} made the interaction between developer and user a seamless process.

GitHub Issues \cite{fiechter2021visualizing} is different from Bugzilla \cite{blei2003latent} and Jira because it does not have a traditional template for reporting a bug. It uses a comment-based system where the user can freely post an issue. The problem with this approach is that it allows users to make a duplicate report and requires a triager to read the report. GitHub Issues \cite{fiechter2021visualizing} addresses this with labels developers can create and attach to the report. The label can be any topic a developer considers appropriate for the issue.

The anatomy of a bug report is broken into product, summary, description, and comments. The product will tell the developer where the user is having issues. The summary will briefly explain the problem, and the description will go more in-depth with details and steps the user tried to fix the issue. The comment section is where the developers and users interact; for GitHub Issues \cite{fiechter2021visualizing}, this is where other users can also comment on a report with suggestions and steps that can potentially fix the bug. The product, summary, description, and comment components of a \textit{bug report} will give our approach the best chance at sorting a bug report into the appropriate category.

\begin{figure}[htp]
    \centering
    \includegraphics[width=8cm]{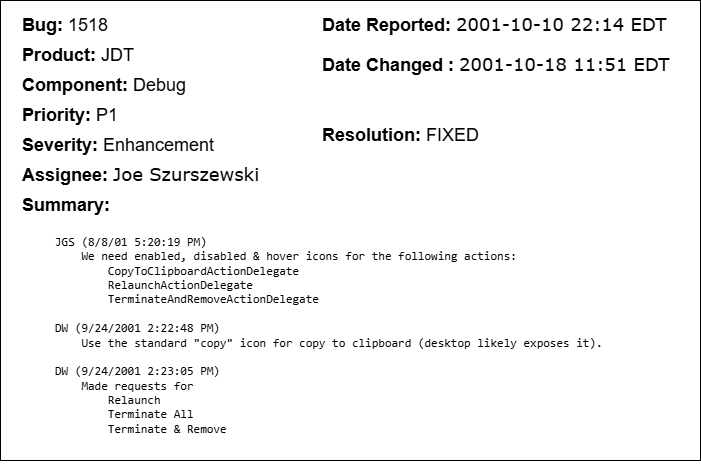}
    \caption{Eclipse bug report \#1518}
    \label{fig: Bug Report Diagram}
\end{figure}

The figure above \ref{fig: Bug Report Diagram} shows an example of Eclipse bug report \#1518; from the \ref{fig: Bug Report Diagram}, we see that the product is the sub-project of Eclipse, and the component is the specific issue the bug was found. Priority and severity are vote-based and set by the trainers and developers, and the summary gives a brief description and example of how to recreate the issue. The date reported is when the bug report was issued, and the date changed is the last date any changes were made to the report. Resolution is a tag that lets developers mark if a bug is fixed, duplicated, or invalid. 

\subsection{Natural Language Processing}
Natural Language Processing (NLP) is a subfield of Artificial Intelligence (AI) \cite{joshi1991natural} that relates to how humans communicate with machines, indulging human language comprehension. NLP uses mathematics and computation to model \textit{lingusitics}. \textit{Lingistics is the scientific study of language significant to NLP with subdivisions including Syntax - the rules of sentence structure; Semantics - the study of linguistic development; and Pragmatics - the analysis of situational context in languages}. The theoretical side of NLP deals with Linguistics and the subdivisions of primary human language. NLP has many concepts that help with various tasks depending on the situation. The concepts we are most interested in are text classification and neural networks. The following sections will briefly introduce language models, text classification, and topic mining. 

\subsubsection{Language Models}
The elementary level of language models is called N-grams. N-grams are probabilistic models that use the words in the sentence to predict the outcome of the next word. The equation is simple: \textit{P} is the probability of the next word, \textit{w} is the word, and \textit{h} is the history (or sequence of previous words in the sentence).
\begin{equation}
    P(w|h)
\end{equation}
Equation 1.\cite{jurafsky2000speech} Probability of word
From equation 1., we can predict the probability of the next word. For example, we can predict the word for the sentence, \textit{The Earth is} by comparing the outcome of \textit{round} or \textit{flat}.
\begin{align}
    P(\text{flat}|\text{The Earth is}) &= 0.2 \\
    P(\text{round}|\text{The Earth is}) &= 0.8
\end{align}
Since the probability of \textit{round} is higher than \textit{flat}, the equation can predict that the following word in the sequence is \textit{round}.

\subsubsection{Topic Mining}
Topic modeling algorithms such as LDA \cite{blei2003latent}, topicMiner MTM \cite{Xia+2017}, and BERTopic \cite{grootendorst2022bertopic} search a whole dataset to group documents into clusters. The traditional approach of topic modeling with LDA \cite{blei2003latent} treats each document as a bag of words approach. LDA \cite{blei2003latent} looks at each word in a document and assigns that word to a topic; the process is repeated until the model has converged documents of a similar nature together. TopicMiner MTM \cite{grootendorst2022bertopic} extends LDA by creating a topic modeling algorithm designed for bug triaging. MTM looks at \textit{bug reports} features instead of summary or description as the central aspect of clustering. The combination of \textit{Product and Component} and the filtered summary text allows MTM to cluster documents based on combinations. BERTopic \cite{grootendorst2022bertopic} is a new topic modeling algorithm; the model uses a combination of traditional and new methods to allow a fully customized model that works for whatever task it is designed for. BERTopic works well with dimensionality and clustering techniques to group documents of similarity into clusters; the model also adds a human representation of the topics to make it easier to understand a topic. 

\subsubsection{Text Classifiers}
Text classifiers read the body of text and assign them to predefined classes. They are suitable for topic modeling because they only look for unique words within the text. The algorithm is designed to \textit{tokenize} a document into the smaller text, removing stop words from the body of text. Stop words appear frequently in a body of text; for example, the, a, and but are some common words. The standard text classification method is supervised machine learning \cite{jurafskyspeech}. Supervised machine learning classification trains on correct data; the data will guide to the correct output. There is a diverse selection of text classification models; the models useful for our approach are Naive Bayes and Neural Network. We will discuss these models more in section \ref{related-work}.

\section{Related Work} \label{related-work}
Automated bug triage, including its sub-problem, automated bug assignment, in open-source software systems has been studied over the past decades (e.g., see \cite{MoinKhansari2010,MoinNeumann2012}). Below, we review related work. In particular, we explain TopicMiner \cite{Xia+2017} and how it categorized each bug report into topics assigned to a developer. We explored LDA \cite{blei2003latent} and explained how we planned to train our model based on the workings of TopicMiner \cite{Xia+2017}. The section broke down the core components of the models and explained how each section was processed. We then explained the concept of Naive Bayes Classifier and Neural Networks. We broke down the core concept of the models and the best use cases for each model.

\subsection{TopicMiner (MTM)} This model extended the Latent Dirichlet Allocation (LDA) \cite{blei2003latent} by converting the words found in bugs into phrases. LDA \cite{blei2003latent} allowed duplicate reports to combine into a single topic and helped bugs that were worded differently separate into their proper topics. TopicMiner \cite{Xia+2017} extended LDA \cite{blei2003latent} by separating each developer into a category that matched the bug reported. Xia et al. \cite{Xia+2017} proposed three phases: model construction, recommendation, and update.

\subsubsection{The model construction phase} During the model construction phase, Xia et al. \cite{Xia+2017} used the training setup from LDA \cite{blei2003latent}. During this stage, MTM \cite{Xia+2017} initialized each word in each \textit{bug report}, and \textit{additional features} were randomly assigned to a topic. The model iterated up to 500 iterations to determine the likelihood estimation of the word belonging to a topic. The process involved three variables: \textit{topic distribution}, \textit{topic assignment vector}, and \textit{topic word vector} \cite{Xia+2017}. MTM calculated the variables using Gibbs Sampling \cite{gelfand2000gibbs}. Gibbs sampling was a Markov Chain Monte Carlo (MCMC) that sampled from the observed word and tried to converge it to the target \cite{jurafsky2000speech}. During the training phase, MTM \cite{Xia+2017} estimated the values of each word in a topic for every bug report by looking for the largest posterior.

\subsubsection{The recommendation phase} In the recommendation phase, TopicMiner \cite{Xia+2017} recommended a list of developers for new bug reports based on reports the individual developers had fixed in the past, as found in the model construction phase. To find the recommended developer for a new bug report, TopicMiner \cite{Xia+2017} combined a bug report's product and component sections into one category and compared the topics to each developer's history of bug fixes. The model then outputs the top 5 recommended developers.

\subsubsection{The model update phase} After MTM \cite{Xia+2017} found the top 5 developers with the highest chance of fixing the new bug report, the model then updated the topic probability of the developer.

\subsection{Latent Dirichlet Allocation (LDA)} Latent Dirichlet Allocation (LDA) \cite{blei2003latent} is a probabilistic model that uses a large body of text such as \textit{documents} and \textit{corpora} to create and distribute topics. LDA \cite{blei2003latent} uses a three-level Bayesian model to classify the topic of a \textit{document}. At the basic level, LDA \cite{blei2003latent} solved the variational inference problem using Bayesian methods. A variational inference approximates the posterior distribution of the latent variables for complex \textit{documents} and \textit{corpora}. Using the variational Expectation Maximization (EM) procedure, LDA \cite{blei2003latent} found an approximate posterior distribution for the model parameters. EM steps are repeated until the E-step finds the optimized value of the variational parameters, and the M-step maximizes the Evidence Lower Bound (ELBO) with respect to the model parameters.

\subsection{Text Classification models} \subsubsection{Naive Bayes Classification} The Naive Bayes Classification \cite{jurafskyspeech} is a simple probability classifier that uses Bayesian Inference to update predictions based on new information. The Naive Bayes Classifier used assumptions to make linear predictions. The first was a bag of words \cite{jurafskyspeech}, which did not consider the word's position when calculating the prediction. The second was the Naive Bayes assumption \cite{jurafskyspeech}, which assumed the word's position.

\subsection{BERTopic} BERTopic \cite{grootendorst2022bertopic} is a transformer-based topic modeling algorithm, unlike traditional methods like LDA \cite{blei2003latent}. BERTopic gave the user complete control over how the model clustered documents together. The algorithm combined techniques used for machine learning algorithms, such as tokenizers, dimensionality reduction, clustering, and embeddings. BERTopic started with document embeddings; the default for this parameter was sentence transformers. The following steps were the most important for our goal. First, we looked at dimensionality reduction; this step was crucial for high-dimensional data and could improve clustering and performance depending on the chosen type. The types of dimensionality reduction we experimented with were Principal Component Analysis (PCA) \cite{mcinnes2020umapuniformmanifoldapproximation} and Uniform Manifold Approximation and Projection for Dimension Reduction (UMAP) 
\cite{mcinnes2020umapuniformmanifoldapproximation}. Once we reduced the dimensions of our dataset, we applied a clustering technique we wanted our model to follow; the default clustering technique was Hierarchical Density-Based Clustering (HDBSCAN) \cite{mcinnes2017hdbscan}. This technique allowed the tuning of parameters to change the metric used for clustering. Once HDBSCAN applied the specified metric, it clustered the documents based on density and metric, removing outliers from the clusters. We also compared HDBSCAN to KMeans \cite{joshi1991natural}. KMeans \cite{joshi1991natural} clustered documents differently; instead of using the density of documents, it set a predefined number of topics to look for and then clustered the documents based on how far apart the document was from the topic. KMeans was helpful since we could set a predefined number of topics. However, we might have encountered issues with developers with a sparse set of bug reports not being represented and developers with a few bug reports unable to cluster \textit{bug reports}. BERTopic used a count vectorizer for the tokenizer by default; we played with this parameter to see if the traditional Term Frequency-Inverse Document Frequency (TF-IDF) was a better option for tokenization. We compared the default settings of BERTopic and tried to fine-tune our parameters to create a cluster of topics for each developer.

\begin{figure}[htp]
    \centering
    \includegraphics[width=8cm]{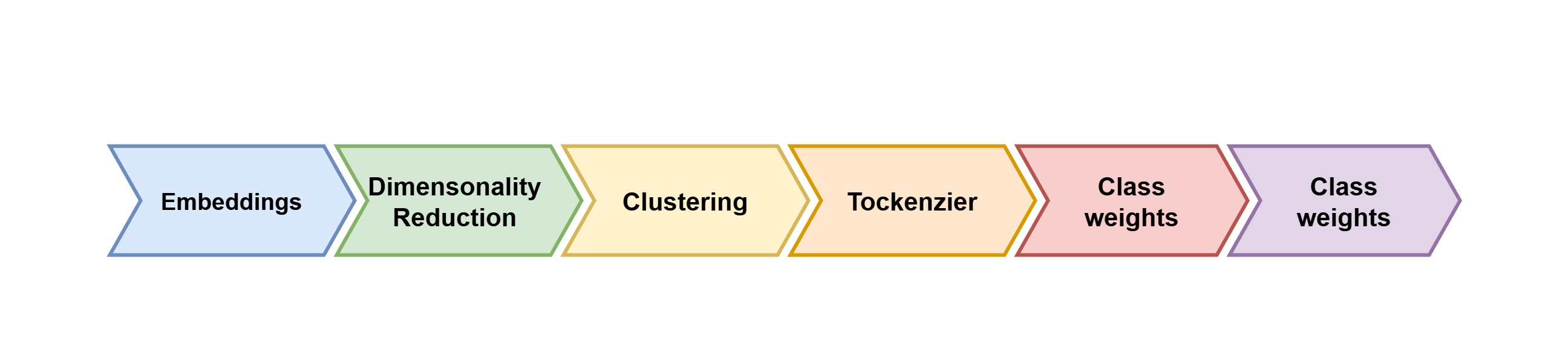}
    \caption{BERTopic Diagram}
    \label{fig: BERTopic Diagram}
\end{figure}

\section{Proposed Approach} \label{proposed-approach}
Based on the work of Xia et al. \cite{Xia+2017}, we will create a topic 
model from the bug report information extracted from unique features. We will then train a text classifier to break the topics further into separate categories. The text classifier will allow our model to find the most capable developers to fix the bug report. Through this approach, we plan to make an open-source and flexible model for future developers to continue the training and verification of accuracy. We propose training an LDA \cite{blei2003latent} on a bug report's features such as \textit{product, component, description, summary}, and \textit{comments}. The topics created from the LDA \cite{blei2003latent} will become the new predefined classes needed for the text classifier. The text classifier is trained on the words pulled from the LDA \cite{blei2003latent}. We will then create a class for developers and assign the number of bugs fixed, a list of bugs fixed in the past, and a ranking system based on the number of bugs fixed for each topic. 
We will use a text classifier to train on each topic and the words associated with the topic. Since traditional text classifiers require classes to be defined before the classification, the topics extracted from the LDA \cite{blei2003latent} will serve as the classes. To accomplish this task, we will experiment with traditional and transformer-based text classifiers and compare and contrast the results of both. 

\begin{table*}[t]
\centering
\caption{Statistics of bug reports}
\begin{adjustbox}{max width=\textwidth}
\begin{tabular}{|l|l|l|l|l|l|l|l|l|}
  \hline
  \textbf{Project} & \textbf{Period} & \textbf{Amount} & \textbf{Product} & \textbf{Component} & \textbf{Combination} & \textbf{Developers} & \textbf{Final Amount} & \textbf{Final Developer} \\
  \hline
  Eclipse & 2001/10/10 - 2013/12/30 & 85,156 & 4 & 21 & 25 & 335 & 42,414 & 296 \\ 
  Mozilla & 1997/03/28 - 2013/12/31 & 205,069 & 13 & 130 & 239 & 2161 & 101,500 & 207 \\ 
  Firefox & 1999/07/30 - 2013/12/31 & 115,814 & 11 & 52 & 119 & 1019 & 19,270 & 36 \\ 
  JDT & 2001/10/10 - 2013/12/31 & 45,296 & 3 & 6 & 11 & 172 & 22,750 & 138 \\ 
  Thunderbird & 2000/04/12 - 2013/12/31 & 32,551 & 1 & 23 & 23 & 380 & 5,703 & 12 \\
    GCC & 1999/01/01 - 2013/09/09 & 129,303 & 4 & 84 & 84 & 272 & 9,971 & 12 \\   
  \hline
\end{tabular}
\end{adjustbox}
\label{tab:Bug_Reports}
\end{table*}

\subsection{Model construction}
We will choose a suitable dataset and handle the preprocessing during the model construction phase. The dataset we use for the initial training is bughub \cite{yuan2024bughub} from the LogPAI team on Git Hub. The dataset comprises nine open-source bug reports from websites like \textit{ Mozilla Firefox, Eclipse Platform, Thunderbird, and Bugzilla}. We will train our topic model to cluster words and sort them into topics. Once sorted, we must optimize each topic to compare word similarity between the topic and text classifier.  

The preprocessing of the dataset will allow us to handle the dataset for topic clustering. We will only focus on \textit{ Title, Description, Summary, and Comments}. We will use the \textit{Components} section to triage bug reports during the training and testing phases. To remove noise, we will omit \textit{ numbers, punctuation, stop words, and variables} from the bug report. Additionally, we need to handle \textit{code snippets} in the reports so that our model is not confused by the complex text form. We will play around with the handle of /textit{code snippets} to see if we can turn them into \textit{bigrams and trigrams} or separate the text altogether. 

After our preprocessing stage, we will create an LDA \cite{blei2003latent} model and pass the clean text bug reports into it in the form of \textit{bag of words (BOW)}. The \textit{bag of words} approach, as talked about in \cite{Xia+2017, ALKHAZI2020106667}, is a way of handling tokenized data from our dataset. \textit{Bag of words} does not consider the word's position; it only looks for the frequency of the word in a \textit{document or bug report}. This allows LDA to cluster the documents using vectorizers like \textit{Term Frequency inverse document frequency (Tf-Idf)} and \textit{Count Vectorizers}. Using one vector over another depends on whether you want to normalize the weight of words that appear more frequently than others.

Once our LDA finishes, we will save the topics and the top \textit{n} words to a new Coma Separated Values (CSV) file to allow the text classifier to train on the topics and learn the vocabulary of words used per topic. We will then test the classifier's accuracy and ability to find the top 1 and 5 prediction accuracies. The top 1 accuracies measure the number of times the classifier identified the correct label, and the top 5 accuracies measure the number of times the text classifier identified the correct label within the top 5 predictions. The top 1 and top 5 accuracies are based on the dataset used by Xia et al. \cite{Xia+2017}. The final stage of our model is the developer classes. This work is based on the research \cite{ALKHAZI2020106667} for ranking developers. We will rank developers based on the number of bugs previously fixed in the past; each developer is placed in a class with the attributes of their \textit{name, weighted list of all bugs, amount of bugs, and top bug fixed}. Once each developer that has fixed over a fixed amount of bugs \cite{Xia+2017} is created, we will store them into a list that will be sorted from most bugs fixed to the most minor bugs fixed. This will allow us to recommend a developer based on the bug report.

Our second approach is to use BERTopic \cite{grootendorst2022bertopic} to create a model for each developer. We will then use a sorting algorithm to filter the developer based on the date of a bug report to test the accuracy and continue training our model for updates. For the models that are over a certain threshold, we will combine clusters and reduce the number of topics; if the model does not have enough valuable information, then we will increase the clustering. We will then save the model and repeat the process. 

\begin{figure}[htp]
    \centering
    \includegraphics[width=6cm]{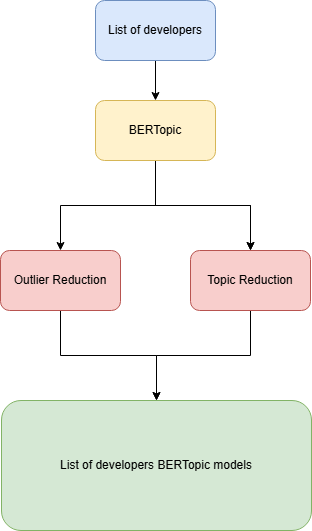}
    \caption{Our Construction Phase}
    \label{fig: Bert construction phase}
\end{figure}

BERTopic \cite{grootendorst2022bertopic} is a valuable tool that allows you to have complete control over the structure of your model. In this subsection, we will break down the creation of the model we used for this approach. The model is constructed together using six individual subsections to guide the model into creating clusters of documents. We started by using transformers to create our document embedding; this was the default for the BERTopic model, and we found that this worked better than traditional embedding methods. We will then use a dimensionality reduction to preserve the information in our data; we will compare Uniform Manifold Approximation and Projection UMAP \cite{mcinnes2020umapuniformmanifoldapproximation} and PCA \cite{wold1987principal} to determine the best. We will compare the use of Hierarchical density-based clustering HDBSCAN \cite{mcinnes2017hdbscan} and KMeans \cite{krishna1999genetic}. Both algorithms work well in different ways. We will use a count vector, remove the stop words from, and experiment with different ngram ranges to determine how increasing or decreasing the ngram range helps with our model. \ref{fig: Bert construction phase} shows a diagram of our model construction. We first train our current developers with over a set amount of bug reports fixed on a shuffled dataset. The dataset is shuffled so that we can train the model to learn the developer's pattern of bug reports to fix. Once the models are trained, we will use our testing data to predict the type of bug report to the developer.

Once the model is trained, we will use a sorting algorithm to filter out the dates and select the active developers during a bug report. We get our inspiration from BT-RL \cite{liu2022automatic} and Avik \cite{Anvik+2006}. They proposed looking at active developers and recommending a new incoming bug report based on the recent activity. We plan to create our model based on the dates a developer was active to test the model's accuracy for predicting a developer based on the time they were active and the new \textit{bug reports}. Priority and severity also play an essential role in the triaging process, so we will add those additional features to the BERTopic model so that the machine can create a profile of the developers based on their underlying topics. Once our algorithm filters developers by the date, priority, and severity level, we will test each BERTopic model that is relevant to an incoming bug report and find the highest probability a developer has of fixing a bug report. We display the \textit{Top-k} number of developers with the best chance of fixing a bug report. If we have an even split of developers who are recommended as the best developer for fixing a \textit{bug report}, we will force the model to choose the best developer based on the date of activity. If a developer activates the most around the time of a \textit{bug report} and the developer has the experience with fixing the type of report, then that developer will become the \textit{Top-1} developer. \ref{} shows the breakdown of the sorting algorithm. 

% ADD Sorting algoithm 

\section{Experimental Results} \label{experimental-results}
We found the \textit{Top-K} accuracy to measure our approach's effectiveness to determine how often our model predicted the correct developer. We did this to compare our results with other research \cite{Anvik+2006}\cite{Xia+2017}\cite{liu2022automatic}. To find the \textit{Top-K accuracy}, we counted the times our model made the correct prediction over the total number of predictions. 

\begin{equation}
    \textit{Top-k} = \frac{\# of correct}{Total \# of predictions}
\end{equation}

We also measured the precision and recall of our model to compare our results with other research papers \cite{Anvik+2006}. To measure recall accuracy, we counted the total number of correct developers over the number of correct developers plus false developers. Precision counted the number of correct developers over the number of correct developers plus the number of incorrect developers.

\begin{equation} 
	\textit{Recall} = \frac{\# \text{ of correct}}{\# \text{ of correct} + \# \text{ of False Negatives}} 
\end{equation}

 \begin{equation} 
	\textit{Precision} = \frac{\# \text{ of correct}}{\# \text{ of correct} + \# \text{ of False Positives}} 
\end{equation}

%\begin{equation}
%    \textit{F1 Score} = 2 \times \frac{\textit{Precision} \times \textit{Recall}}{\textit{Precision} + \textit{Recall}}
%\end{equation}

\subsection{Experimental Setup}
We used 486,591 bug reports from the BugHub GitHub repository \cite{yuan2024bughub}, which utilized bug reports from Bugzilla \cite{1407819} websites. We added three additional columns to each bug report: \textit{Assignee Real Name} for the name of the Developer, \textit{Product} to indicate the platform where the bug was occurring, and \textit{Severity} to denote how severe the bug report was. With this new \ textit {features}, we created a BERTopic model \cite{grootendorst2022bertopic} for each Developer relevant to the Project. We filtered our dataset based on previous works \cite{Xia+2017}\cite{Anvik+2006}\cite{liu2022automatic} and determined a developer’s relevance to a project based on the number of bugs fixed in the past. We further filtered the dataset by removing generic developer names, as shown in \cite{Xia+2017}, and provided a breakdown of the final number of developers. The \textit{Project} referred to the name of the bug reports and their source, \textit{Amount} was the total number of bug reports in the dataset, \textit{Product} indicated the total number of products in the bug reports, \textit{Component} referred to the total number of components, \textit{Combination} was the total number of combinations of Product and Component, and \textit{Developer} was the total number of developers. Additionally, we filtered out developers who had fixed over 100 \textit{bug reports}, which helped the BERTopic model create enough clusters to predict \textit{bug reports}. If we applied different dimensional reduction and clustering methods, developers who had fixed fewer than 100 \textit{bugs} were considered.

Each developer had a BERTopic model fine-tuned to the \textit{bug reports} marked as \textit{fixed} by that developer to ensure the models received accurate information about the developer's past. We first filtered our dataset by the developer and then performed an 80/20 split on the developer's bug list. Once the model was trained, we reduced the number of topics for larger models to a smaller size; we tested different values and found that reducing the more prominent models to ten topics allowed them to retain a high probability score while giving smaller models a chance to be a top pick. We reduced the number of outliers for smaller models, which allowed the models to have more representation from the developer's documents. Once a developer's BERTopic model was complete, we saved the model and repeated the process for the remaining developers. 

\subsection{Results}
We applied a simple accuracy test to our model and compared it to TopicMiner MTM \cite{Xia+2017}, BT-RL \cite{liu2022automatic}, LDA-SVM, LDA-KL, and BUGZIE \cite{tamrawi2011fuzzy}. Among the baseline models, TopicMiner MTM \cite{Xia+2017} performed the best compared to the other approaches. We compared Eclipse, Mozilla, Firefox, Thunderbird, and JDT bug reports. We measured the \textit{Top-K} accuracy of our model, focusing on \textit{Top-1} through \textit{Top-5} accuracies. We treated the testing phase as the recommendation phase for a new bug report received, so we needed to ensure that our model was trained on the datasets in a time sequence manner. Each developer's bug reports were split into a test and training set, meaning the models will predict new bug reports unseen by the developer up to the time report split. This approach gave us a real-world view of how accurately our model predicted the correct developer. 

% Table for Top-1 Accuracy
\begin{table}[h]
    \centering
    \caption{Top-1 Accuracy}
    \resizebox{\columnwidth}{!}{%
    \begin{tabular}{|l|l|l|l|l|l|}
    \hline
        \textbf{Projects} & \textbf{BERTopic} & \textbf{MTM} & \textbf{BZ} & \textbf{BT-RL} & \textbf{LDA-SVM} \\
        \hline
         Eclipse & - & 0.64 & 0.39 & 0.46 & 0.15\\ 
         Mozilla & - & 0.52 & 0.30 & 0.20 & 0.15\\ 
         Firefox & - & - & - & - & -\\ 
         JDT & 0.97 & - & - & - & -\\ 
         Thunderbird & 0.82 & - & - & - & -\\
         GCC & 0.88 & 0.51 & 0.27 & - & 0.21 \\
         \textbf{Average} & 0.89 & 0.56 & 0.32 & 0.33 & 0.17\\
         \hline
    \end{tabular}%
    }
    \label{Top1Accuracy}
\end{table}

% Table for Top-5 Accuracy
\begin{table}[h]
    \centering
    \caption{Top-5 Accuracy}
    \resizebox{\columnwidth}{!}{%
    \begin{tabular}{|l|l|l|l|l|l|}
    \hline
        \textbf{Projects} & \textbf{BERTopic} & \textbf{MTM} & \textbf{BZ} & \textbf{BT-RL} & \textbf{LDA-SVM} \\
        \hline
         Eclipse & - & 0.90 & 0.71 & 0.78 & 0.35\\ 
         Mozilla & - & 0.78 & 0.72 & 0.68 & 0.32\\ 
         Firefox & - & - & - & - & -\\ 
         JDT & 1.00 & - & - & - & -\\ 
         Thunderbird & 0.92 & - & - & - & -\\
         GCC & 0.91 & 0.80 & 0.59 & - & 0.49 \\
         \textbf{Average} & 0.94 & 0.83 & 0.67 & 0.73 & 0.39\\
         \hline
    \end{tabular}%
    }
    \label{Top5Accuracy}
\end{table}

% Need to finish getting values, and talk about results for this section

% Need to add precision, recall, and f1-score accuracy to the paper 

% Maybe try time seris approach as well

% Need training time and testing time?

\section{Discussion} \label{discussion}

\subsection{Research Questions}

\subsubsection{\textbf{RQ1.What impact do different parameter settings have on the quality of the topics generated by BERTopic?}}
We tested the dimensionality reduction and clustering parameters for BERTopic since they are the most important for our case. With dimensionality reduction, we compared PCA \cite{wold1987principal} and UMAP \cite{mcinnes2020umapuniformmanifoldapproximation}. We found that we could interchange the two and have no significant impact on the effects of our clustering. We chose to keep UMAP\cite{mcinnes2020umapuniformmanifoldapproximation} since it is flexible and allows more control over the reduction. For clustering, we compared KMeans \cite{krishna1999genetic} and HDBSCAN \cite{mcinnes2017hdbscan}. We found that HDBSCAN worked the best for all developers since it uses outliers and does not require a predefined number of clusters. This was useful to developers who have a few bugs fixed. 

\subsubsection{\textbf{RQ2.What are the computational resource requirements for training and using BERTopic, and how do they scale with dataset size?}}
Since BERTopic \cite{grootendorst2022bertopic} uses sentence transformers \cite{reimers-2019-sentence-bert} with the PyTorch library, a Graphical Processing Unit (GPU) is the ideal way of using BERTopic. You can still use a Central Processing Unit (CPU), but processing may take longer. The dataset used for training only depends on the number of \textit{bug reports} and the number of developers each BERTopic \cite{grootendorst2022bertopic} model needs to train. 

\subsubsection{\textbf{RQ3.How difficult is it to implement this approach into an existing project?}}
Implementing our approach requires some setup for an existing project to get the best results. A dataset with all of the relevant developers and the features is required. The features should include \textit{product, component, priority, severity, and a description} of the bug reports developers have fixed. Additionally, an \textit{open or creation date} is beneficial for finding the active times of each developer for recommendations of future bugs.

\subsubsection{\textbf{RQ4.What is the ideal scenario for a model like this?}}
The ideal scenario of our approach is recommending the best or \textit{Top-K} best developers that can fix an incoming \textit{bug report}. We also have found that for bugs marked as \textit{unassigned} in the past with no fixer, our model can find a recommendation for a developer using the times when a developer is active. The model will recommend a developer based on the amount and types of \textit{bug reports} assigned to a developer during a \textit{unassigned} report. 

\subsection{Threats to Validity} \label{threats-to-validity}
\subsubsection{Internal Threats}
We tried to find and recreate the models we used to compare our results but could not find them. So, we compared the results of the models obtained in their research with the appropriate project to keep the measurements and compare our model similarly. Our approach can take up a lot of space depending on the number of developers since we are creating one BERTopic \cite{grootendorst2022bertopic} model per each developer. Additionally, adding a new developer is easy however for our model to best work it is recommended that the developer has fixed over a certain amount of \textit{bug reports} before creating a model. The issue can be fixed with using a different type of clustering and dimensional reduction, we found that KMeans \cite{krishna1999genetic} and PCA \cite{wold1987principal} are good options to start and test the difference. Our approach also relies on \textit{bug reports} with dates as we found the best for filtering developers.

\subsubsection{External Threats}
The \textit{bug reports} we used for our testing was pulled from the bughub GitHub \cite{yuan2024bughub}. The total amount of bug reports used from the repository totaled 486,591 reports. The projects used for the research were limited to sub-projects of Eclipse and Mozilla. The Eclipse sub-projects are Eclipse-platform and Eclipse-JDT. The Mozilla sub-projects are Mozilla-firefox, Mozilla-core, and Mozilla-Thunderbird. Since the dataset used with our model is limited, we need to apply different projects so that we can get a broader range of different types of bug reports. 

\section{Conclusion and Future Work} \label{conclusion-future-work}
We made the model for this project open-source and encouraged the development of our project for future research and projects. In the future, we plan to improve the model's efficiency to see if using the BERTopic \cite{grootendorst2022bertopic} model merge feature significantly impacts the predictability of new bug reports to developers in the future. We would also like to improve our filtering process using a neural network or a decision tree. That way, we can train the model on the dates of a developer and allow it to filter the developers automatically. We will also add more projects and bug reports to test our model from \textit{bug list} on Bugzilla's website, allowing us to test a broader range of developments and developers. The goal is to be able to apply the model with something like GitHub issues \cite{fiechter2021visualizing}; as of now, we will continue to use Bugzilla for its ease of access. 

We applied a traditional approach to automatic bug triaging using topic modeling. We get our motivation from the works of previous research papers \cite{blei2003latent}\cite{Anvik+2006}\cite{Xia+2017}\cite{liu2022automatic} and extend the works of automatic bug triaging with a new approach of applying BERTopic \cite{grootendorst2022bertopic} and using a sorting algorithm. We differ from the previous research by creating one topic model per developer, allowing us to cluster the \textit{bug reports} marked as \textit{FIXED} to create a pattern of \textit{bugs} the developer has the best chance of fixing. We also found that we can filter our developers using the dates, priority, and severity of \textit{bug reports} to allow our model a better chance of creating a \textit{Top-K} of developers with the best experience to fix a particular bug. From this, we can guide our model to look at developers and choose who is best at fixing a certain \textit{bug report} based on the experience level or the number of years they actively worked on a project. With the additional features, we added to the BERTopic models, and the sorting algorithm applied to developers once a new \textit{bug report} is received, we were able to beat previous research papers by a significant amount. We can further improve the accuracy of our model by applying a Neural Network or some other kind of decision-based algorithm and learning the developer's level of experience based on the total number of years, bug reports, priority, and severity level bug reports they have fixed. Adding other types of models trained on years of experience with the ability to look at the type and time frame of a bug report would help the model learn how to categorize in a history-based instead of only looking at the features of a \textit{bug report} we learned that we could help our model make correct predictions through the use a sorting algorithm. The sorting algorithm with BERTopic \cite{grootendorst2022bertopic} opens the door for future work to improve the efficiency and learning of the model.

\section*{Software and Data Availability}
The prototype is available under a permissive open-source license at \url{https://github.com/qas-lab/MarshallReu}. The data used for the evaluation are also publicly available at \url{https://github.com/logpai/bughub/tree/master/EclipsePlatform}.

\section*{Acknowledgment}
This material is based upon work supported by the U.S. National Science Foundation (NSF) under Grant No. 2349452. Any opinions, findings, conclusions, or recommendations expressed in this material are those of the authors and do not necessarily reflect the views of the NSF.

\bibliography{refs}
\bibliographystyle{IEEEtran}

\end{document}